\RequirePackage{fix-cm}
\documentclass[twocolumn,epjc3]{svjour3}  
\usepackage{amsmath}
\usepackage{amssymb}
\smartqed  
\RequirePackage{graphicx}
\usepackage{hyperref}
\hypersetup{colorlinks, linkcolor=blue,  citecolor=blue,  urlcolor=blue}
\journalname{Eur. Phys. J. C}

\begin{document}
\title{Accelerating Kerr solution in SU(1,1)/U(1)-sigma model }

\author{Wanke Hu\thanks{email: frank-huwanke@outlook.com}}

\institute{School of Physics and Astronomy, China West Normal University, Nanchong, Sichuan 637002, People's Republic of China}

\date{Received: date / Accepted: date}
\maketitle

\begin{abstract}
Under the coset formulation of pure Einstein spacetime, we solve the reduction problem of the nonlinear \(\sigma\)-model for this spacetime and present the process of the inverse scattering technique after simplification. Through the simplified inverse scattering process, taking the pure Rindler metric as the background, we introduce a pair of solitons that are complex conjugates of each other and obtain an accelerating solution in the form of three solitons in this model. By means of appropriate parameter selection and coordinate transformation, we prove that this accelerating solution in the three-soliton formulation is equivalent to the accelerating Kerr solution.
\end{abstract}

\section{Introduction}\label{intro}

For a long time, due to the fact that the Einstein field equations are multivariate, high-order, and nonlinear equations, which are extremely complex, the search for exact solutions of the Einstein field equations has been greatly restricted. Among numerous solution generation techniques, the inverse scattering technique proposed by Belinski and Zakharov(BZ) is undoubtedly very effective \cite{ref1,ref2}. This method requires rewriting the system into an integrable system, which depends on the symmetry conditions of the system. To be precise, in four-dimensional gravity theory, the characteristic symmetry group in the vacuum gravitational field is \(\text{SL}(2,\mathbb{R})\) \cite{ref3}, the Einstein-Maxwell theory corresponds to \(\text{SU}(2,1)\) \cite{ref4,ref5}, and the characteristic symmetry group in the Einstein-Maxwell-dilaton-axion theory (EMDA) is \(\text{Sp}(4,\mathbb{R})/\text{U}(2)\)  \cite{ref6,ref7}. Such systems with characteristic symmetry groups are called nonlinear \(\sigma\)-model. The coupling situation of the gravitational field and the electromagnetic field has also been constructed in Ref. \cite{ref8}, and for details, one can also refer to Refs. \cite{ref9,ref10}.

The inverse scattering technique proposed by BZ is carried out in the coordinate representation space. This method requires a known solution as the background, which is called the background solution or seed solution. New solutions are obtained by adding different numbers of solitons to this background solution. This paper mainly introduces another type of coset representation. The coset representation of the vacuum Einstein spacetime corresponds to the \(\text{SU}(1,1)/\text{U}(1)\)-sigma model \cite{ref11}. In addition, the coset representation of the Einstein-Maxwell spacetime of \(\text{SU}(2,1)/(\text{SU}(2)\times\text{U}(1))\) is also given in Ref. \cite{ref12}. When applying the inverse scattering technique in the nonlinear \(\sigma\)-model, it is necessary to use the characteristic symmetry group of the system to solve the reduction problem, which leads to the solitons appearing in pairs as complex conjugates on the imaginary axis. It has been given in Ref. \cite{ref13} that by adding a pair of complex solitons with the Minkowski spacetime as the background, under the coset representations of pure gravity and Einstein-Maxwell respectively, the soliton solutions of the Kerr-NUT-like and Kerr-Newman-like are given. And in the coordinate representation, Ref. \cite{ref14} takes the pure Rindler metric as the background and adds a pair of solitons located on the real axis to obtain the soliton solution of the accelerated Kerr-NUT metric. However, in the nonlinear \(\sigma\)-model, there has been no work on using the inverse scattering technique to obtain accelerating solutions. The main purpose of this paper is to use the inverse scattering technique to derive accelerating solutions under the coset formulation of pure Einstein spacetime. This is merely a warm-up, aiming to provide ideas for the reduction problem of the Einstein-Maxwell theory and the EMDA theory's nonlinear \(\sigma\)-model, so as to derive accelerating solutions in these two theories. 

In this paper, under the coset representation of the pure Einstein theory, we solve the reduction problem of two solitons. By introducing a pair of virtual solitons, we obtain the soliton solution of the extremely accelerated Kerr metric in this nonlinear \(\sigma\)-model for the first time. And we prove that through appropriate coordinate transformations and analytic continuations, this soliton solution can be converted into the commonly formed accelerated Kerr solution \cite{ref15}. In Scet. \ref{sec2}, we will briefly review the inverse scattering theory in the nonlinear \(\sigma\)-model and give the reduction conditions in this model, thus simplifying the steps of generating soliton solutions. In Scet. \ref{sec3}, we will take the pure Rindler metric as the background, introduce a pair of complex solitons, and give the form of the soliton solution. Then we will prove that this soliton solution is the accelerated Kerr solution through appropriate parameter selection and coordinate transformation. In Scet. \ref{sec4}, it is the conclusion and discussion of this paper. Regarding the process of using the characteristic symmetry group of the nonlinear \(\sigma\)-model to solve the reduction problem, we put it in \ref{appendix}, and readers who are interested can read it.

\section{Inverse scattering theory in nonlinear \texorpdfstring{$\sigma$}{sigma}-model}\label{sec2}

In this section, we will briefly review the inverse scattering method within the nonlinear
SU(1,1)/U(1) $\sigma$-model of the four-dimensional vacuum gravity theory.

In a stationary axisymmetric spacetime (SAS) where exists two Killing vectors: one time-like
$\partial_t$ and another space-like $\partial_\phi$, its line element can be written in the Weyl-Papapetrou
metric form:
\begin{equation}\label{wp}
    ds^2 = -f(dt - \omega d\phi)^2 + \left[ \frac{e^{2\gamma}}{f} (dz^2 + d\rho^2) +\frac{\rho^2}{f}  d\phi^2 \right],
\end{equation}
where \( f \), \( \omega \), and \( \gamma \) are all functions of \( \rho \) and \( z \). It is pointed out in Ref. \cite{ref2} that once we know \( f \) and \( \omega \), the function \( \gamma \) can be determined by integration. Then the Einstein field equations are simplified to Ref. \cite{ref16}:
\begin{equation}\label{eq1}
    \begin{split}
& f \left( \partial_\rho^2 + \rho^{-1} \partial_\rho + \partial_z^2 \right) f - (\partial_\rho f)^2 - (\partial_z f)^2 \\
&  \quad+ \left( \rho^{-1} f^2 \partial_\rho \omega \right)^2 + \left( \rho^{-1} f^2 \partial_z \omega \right)^2 = 0, \\
& \partial_\rho \left( \rho^{-1} f^2 \partial_\rho \omega \right) + \partial_z \left( \rho^{-1} f^2 \partial_z \omega \right) = 0,
\end{split}
\end{equation}
for this system of equations, a twist potential \( \psi \) can be introduced, satisfying the following relations:
\begin{equation}\label{cosettran}
    \partial_\rho \psi = \rho^{-1} f^2 \partial_z \omega, \quad \partial_z \psi = -\rho^{-1} f^2 \partial_\rho\omega,
\end{equation}
the system of equations can be rewritten as:
\begin{equation}\label{eq2}
    \begin{split}
& f \left( \partial_\rho^2 + \rho^{-1} \partial_\rho + \partial_z^2 \right) f - (\partial_\rho f)^2 - (\partial_z f)^2 \\
&\quad+ (\partial_\rho \psi)^2 + (\partial_z \psi)^2 = 0, \\
& \partial_\rho \left( \rho f^{-2} \partial_\rho \psi \right) + \partial_z \left( \rho f^{-2} \partial_z \psi \right) = 0,
\end{split}
\end{equation}
to express it in a more concise form:
\begin{equation}
    \partial_\rho \left( \rho \partial_\rho \mathbf{P} \cdot \mathbf{P}^{-1} \right) + \partial_z \left( \rho \partial_z \mathbf{P} \cdot \mathbf{P} \right) = 0,
\end{equation}
where
\begin{equation}\label{P1}
   \mathbf{P} = \frac{1}{f} \begin{pmatrix} 1 & \psi \\ \psi & f^2 + \psi^2 \end{pmatrix},
\end{equation}
this is equivalent to being expressed by the Ernst potential \( \varepsilon = f + i\psi \):
\begin{equation}\label{P2}
    \mathbf{P} = \frac{1}{\bar{\varepsilon} + {\varepsilon}} 
    \begin{pmatrix}
    2 & i(\bar{\varepsilon} - \varepsilon) \\ i (\bar{\varepsilon}- \varepsilon) &2 \bar{\varepsilon} \varepsilon
    \end{pmatrix}.
\end{equation}
satisfies the following conditions:
\begin{equation} \label{conP}
    \mathbf{P} = \mathbf{P}^{\dagger}, \quad (\Lambda \mathbf{P})^2 = \mathbf{I} , \quad \det(\mathbf{P}) = 1 ,
\end{equation}
where
\begin{equation}\label{conGamma}
    \Lambda = \begin{pmatrix} 
0 & i \\
-i & 0 
\end{pmatrix},
\end{equation}
Note that dagger denotes the Hermitian conjugate, \(\mathbf{I}\) represents the identity matrix and its order is the same as that of matrix \(\mathbf{P}\). 

The vacuum Einstein field equation in SAS is written as:
\begin{subequations}
\begin{equation}\label{E1}
\nabla \cdot \left[\rho (\nabla \mathbf{P}) \mathbf{P}^{-1}\right] = 0,
\end{equation}
\begin{equation}\label{E2}
(\gamma)_{,\rho} = \frac{1}{8\rho}\operatorname{Tr}\left[\mathbf{U}^2 - \mathbf{V}^2\right],
\end{equation}
\begin{equation}\label{E3}
(\gamma)_{,z} = \frac{1}{4\rho}\operatorname{Tr}(\mathbf{UV}),
\end{equation}
\end{subequations}
where
\begin{equation}\label{UV}
    \mathbf{U} = \rho (\partial_{\rho} \mathbf{P}) \mathbf{P}^{-1}, \quad \mathbf{V} = \rho (\partial_{z} \mathbf{P}) \mathbf{P}^{-1}.
\end{equation}

Once we know the matrix \(\mathbf{P}\), we can obtain the line element of spacetime through integrating Eqs. (\ref{E2}) and (\ref{E3}). The complete integrability of Eq. (\ref{E1}) can be expressed by the compatibility condition of the Lax pair with a \(2\times2\) generating matrix \( \boldsymbol{\Psi}(\lambda, \rho, z) \):
\begin{equation}\label{LA}
    \begin{cases}
D_\rho\boldsymbol{\Psi} \equiv \partial_{\rho} \boldsymbol{\Psi} + \dfrac{2\lambda\rho}{\lambda^2 + \rho^2} \partial_{\lambda} \boldsymbol{\Psi} = \dfrac{\rho \mathbf{U} + \lambda \mathbf{V}}{\lambda^2 + \rho^2} \boldsymbol{\Psi}, \\
D_z\boldsymbol{\Psi} \equiv \partial_{z} \boldsymbol{\Psi} - \dfrac{2\lambda^2}{\lambda^2 + \rho^2} \partial_{\lambda} \boldsymbol{\Psi} = \dfrac{\rho \mathbf{V} - \lambda \mathbf{U}}{\lambda^2 + \rho^2} \boldsymbol{\Psi}.
    \end{cases}
\end{equation}

In the theory of solitons and inverse scattering, we can obtain different new solutions by adding solitons to a known background metric (also called a seed solution). More precisely, if we use the inverse scattering technique to add solitons to a given specific background \(\mathbf{P_0}\), we can obtain any matrix \(\mathbf{P}\) that satisfies the system of Eqs. (\ref{E1}-\ref{E3}) .

Firstly, we briefly introduce the inverse scattering steps in the spacetime after this duality. Since the transmission amplitude is only a part of the eigenfunction \( \boldsymbol{\Psi}(\lambda, \rho, z) \), this function has the same single pole structure in some arbitrarily large but finite region in the \( \lambda \) plane. \(\lambda\) is a complex spectral parameter independent of \(\rho\) and \(z\). For convenience, hereinafter we will abbreviate \(\boldsymbol{\Psi}(\lambda, \rho, z)\) as \(\boldsymbol{\Psi}\). Therefore, \(\boldsymbol{\Psi}\) and \(\boldsymbol{\Psi_0}\) can have the following relationship:
\begin{equation}\label{psi=chipsi0}
 \boldsymbol{\Psi}=\chi\boldsymbol{\Psi_0}   ,
\end{equation}
the dressing matrix \(\chi\) is a matrix function of \(\lambda\), \(\rho\), and z, and has the following form:
\begin{equation}\label{chi}
\chi(\lambda)=\mathbf{I} + \sum_{k = 1}^{N}\frac{R_k}{\lambda-\mu_k} ,
\end{equation}
we can define the inverse of \( \chi \) as:
\begin{equation}\label{inchi}
\chi^{-1}(\lambda)=\mathbf{I} +\sum_{k = 1}^{N}\frac{S_k}{\lambda-\nu_k} ,
\end{equation}
among them, \( R_k \) and \( S_k \) are matrices with respect to \( \rho \) and \( z \), and their number of orders is the same as that of the background metric \( \mathbf{P_0} \). \( \mu_k \) and \( \nu_k \)  depend only on \( \rho \) and \( z \). The pole \( \mu_k \) represents a soliton. No physical meaning is assigned to \(\nu_k\) for the time being, instead, it is only defined as a first-order pole in mathematics. And \( N \) represents the number of poles. Theoretically, the order of the poles can be arbitrary, but for the sake of simplicity, only first-order poles are considered in this paper.

When \(\lambda = 0\), it can be seen that Eq. (\ref{psi=chipsi0}) are the same as Eq. (\ref{E1}), so we have the following:
\begin{equation}\label{P=chiP0}
   \mathbf{P} = \chi(0) \mathbf{P_0}.
\end{equation}

Obviously, after we present a coset formulation \( \mathbf{P_0} \) for the background metric, if we can find the expression for \( \chi(0) \), we can directly compute the new solution \( \mathbf{P} \) after adding a soliton. Finding the expression for \( \chi(0) \) will make use of the constraint problem in the nonlinear sigma model. For simplicity, we directly present the key formula, and the detailed derivation steps will be given in the appendix.
\begin{equation}\label{RS}
    R_k=n_k m_k^\dagger,\quad S_k=p_k q_k^\dagger ,
\end{equation}
where
\begin{equation}\label{nm}
\begin{aligned}
   &m_k= m_{0k}\mathbf{\Psi_0}^{-1}(\mu_k),\quad p_k=\mathbf{\Psi_0}(\nu_k)p_{0k},\\
    &n_k = \sum_{l=1}^{N} (\Gamma^{-1})_{kl} \frac{\mathbf{P_0} \Lambda p_l}{{\nu_l}}, \quad q_k = \sum_{l=1}^{N} (\Gamma^{-1})_{lk} \frac{\Lambda\mathbf{P_0} m_l}{{\mu_l}},\\
    & \Gamma_{kl} = \frac{p_l^{\dagger} \Lambda \mathbf{P_0} m_k}{\rho^2 +{\mu}_k \nu_l} .
\end{aligned}
\end{equation}
Among them, \( m_{0k} \) and \( p_{0k} \) are arbitrary complex vector with two components, and bar denotes the complex conjugate.

To ensure that the solution \( \mathbf{P} \) we obtain also belongs to the \( \text{SU}(1,1)/\text{U}(1) \) sigma-model, by solving two reduction problems (the detailed derivation is given in the appendix), we also need the following additional conditions:
\begin{equation}\label{reduction}
    \begin{aligned}
        &\mu_i \mu_{i+1}=\nu_i \nu_{i+1}=-\rho^2,\quad (i=1,2,..,N-1)\\
        &m_{i+1}=\Lambda \mathbf{P_0} m_i, \quad p_{i+1}= \mathbf{P_0} \Lambda p_i,\\
        &p_k=\Lambda m_k, \quad \bar{\nu_k}=\mu_k, .
    \end{aligned}
\end{equation}

Thus, we only need to assume two arbitrary complex parameters in \( m_{0k} \) and \( p_{0k} \), then calculate Eqs. (\ref{nm}) and (\ref{RS}) respectively, and substitute them into Eq. (\ref{P=chiP0}) to obtain the new solution \( \mathbf{P} \).

\section{Accelerating Kerr solution in nonlinear \texorpdfstring{$\sigma$}{sigma}-model}\label{sec3}

In this chapter, we will take the Rindler metric as the background. This background solution has been thoroughly explained and verified, and the relevant references are provided in Refs. \cite{ref14}\cite{ref23}\cite{ref24}. By using the inverse scattering technique, we will add a pair of solitons located on the imaginary axis, and present a metric with a 3 - soliton formulation. Then, by determining the parameters and performing coordinate transformations, we will prove that this solution is the accelerating Kerr metric. The Rindler metric in the form of a single soliton can be obtained by adding a single soliton in flat space. It should be noted that if we want to obtain the Rindler metric with the Lorentz signature, it is necessary to take the Euclidean flat spacetime as the background \cite{ref14}\cite{ref17}. Here, we directly present: 
\begin{equation}\label{Rindler}
    ds^2 = -\mu_3 dt^2 + \frac{\rho^2}{\mu_3} d\phi^2 + \frac{\mu_3}{\rho^2 + \mu_3^2}(d\rho^2 + dz^2),
\end{equation}
among them, \( \mu_3 \) is the soliton, and the subscript 3 is only used to distinguish it from the two solitons we add with this metric as the background. We can write the background \( \mathbf{P_0} \):
\begin{equation}\label{P0}
   \mathbf{P_0 } = \begin{pmatrix} 
\frac{1}{\mu_3} & 0 \\
0 & \mu_3 
\end{pmatrix},
\end{equation}
substituting Eq. (\ref{P0}) into Eqs. (\ref{UV}) and (\ref{LA}), we can obtain \( \mathbf{\Psi_0} \): 
\begin{equation}\label{psi0}
\mathbf{\Psi_0} = \begin{pmatrix} 
\frac{1}{\mu_3- \lambda} & 0 \\
0 & \mu_3 - \lambda
\end{pmatrix},
\end{equation}

Next, we can define the vector \( m_{01} \) with arbitrary complex parameters.  Therefore, in the following text, we will use \( \mu_1 \to \mu \) to replace them. The following is:
\begin{equation}\label{n0}
    m_{01} = (c_1, c_2),
\end{equation}
where \( c_1\) and \(c_2\) are an arbitrary complex number, and substituting \( m_{01} \) and \( \mathbf{\Psi_0} \) into Eqs. (\ref{nm}) and (\ref{reduction}), we can write:
\begin{equation}\label{mk}
    m = \begin{pmatrix} -c_1 \mu_{23}, & \frac{-c_2}{\mu_{23}}  \\
    \frac{-i c_2 \mu_3}{\mu_{23}}, &\frac{i c_1 \mu_{23}}{\mu_3}
    \end{pmatrix},
\end{equation}\\
\begin{equation}\label{pk}
    p = \begin{pmatrix}
\frac{i c_2}{\mu_{23}} & -i c_1 \mu_{23} \\
\frac{c_1 \mu_{23}}{\mu_3} & \frac{\mu_3 c_2}{\mu_{23}}
\end{pmatrix}.
\end{equation}

We introduced the following notations:
\begin{equation}
    R_{kl}=\rho^2+\mu_k \mu_l, \quad \mu_{kl}=\mu_k-\mu_l,
\end{equation}
it should be noted that here, \( \mu_1 \to \mu \), \( \mu_2 \to \overline{\mu} \), and \( \mu_3 \to \mu_3 \).

The components of the Gamma matrices are:
\begin{equation}\label{Gamma}
\begin{aligned}
\Gamma_{11} &= \frac{\overline{c_1} c_1 \mu_{13}^2 \mu_{23}^2 + \overline{c_2} c_2\mu_3^2} {\mu_3 \mu_{23} \mu_{13} R_{12}}, \\
\Gamma_{12} &= \frac{i(-\overline{c_1 }c_2\mu_{13}^2 + c_1\overline{c_2} \mu_{23}^2) \mu}{\mu_{13} \mu_{23} \rho^2 \mu_{12}}, \\
\Gamma_{21} &= \frac{-i(-\overline{c_1 }c_2\mu_{13}^2  + c_1\overline{c_2} \mu_{23}^2) \overline{\mu}}{\mu_{23} \mu_{13} \rho^2 \mu_{12}}\\
\Gamma_{22} &= -\frac{(\overline{c_1} c_1\mu_{13}^2 \mu_{23}^2 +  \overline{c_2} c_2\mu_3^2) \mu \overline{\mu}}{\mu_3 \mu_{13} \mu_{23} \rho^2 R_{12}}, 
\end{aligned}
\end{equation}

Finally, using Eqs. (\ref{P=chiP0}) and (\ref{RS}), we calculate the new solution \( \mathbf{P} \) in the form of a three-soliton. Clearly, the matrix \( \mathbf{P} \) we obtained naturally satisfies condition (\ref{conP}). It should be noted that our method can only add an even number of solitons to the background solution, if one wishes to add an arbitrary number of solitons, some conditions need to be removed (detailed steps will be provided in the appendix). After obtaining the new solution \( \mathbf{P} \), we can write:
\begin{equation}\label{f1}
\begin{aligned}
f &= \mu \bar{\mu} \mu_3 \frac{A}{B},\quad 
\psi = \frac{D}{H},\quad e^{2\gamma} = C_0 \mu \bar{\mu} \mu_3\frac{A}{H},
\end{aligned}
\end{equation}
among them,
\begin{equation}\label{A1}
\begin{aligned}
A =& -\mu_{12}^2 \left(c_1 \overline{c_1}\mu_{23}^2 \mu_{13}^2 + \overline{c_2} c_2\mu_3^2\right)^2 \rho^2 \\
  &+\mu_3^2 \left(-\overline{c_1} c_2\mu_{13}^2  + \overline{c_2} c_1 \mu_{23}^2\right)^2 R_{12}^2 ,\\
B =& \mu_{23}^4 \left( \rho^4 \mu_{12}^2 \overline{c_1}^2 \mu_{13}^4 + \overline{c_2}^2 R_{12}^2 \mu_3^2 \mu^2 \right) c_1^2 \\
  &- 2 c_1 \overline{c_1} c_2 \overline{c_2} \mu_{23}^2 \mu_{13}^2 \mu_3^2 \mu^2 \overline{\mu} R_{11} R_{22}\\ 
  &+ c_2^2 \overline{\mu}^2 \mu_3^2 \left( \overline{c_2}^2 \mu_3^2 \mu^2 \mu_{12}^2 + R_{12}^2 \overline{c_1}^2 \mu_{13}^4 \right) ,\\
D =& \mu_{12}\mu_3^2 R_{12} \left( -\overline{c_1 c_2\mu} \mu_{13}^2  R_{11} \left( \mu_{23}^4 c_1^2 + \mu_3^2 c_2^2 \right) \right.
\\&\left.  + c_1 c_2 \mu_{23}^2 \mu R_{22} \left( \overline{c_1}^2 \mu_{13}^4 + \overline{c_2}^2 \mu_3^2 \right) \right) ,\\
H =&  \mu \bar{\mu} \mu_3 \mu_{13}^2 \mu_{23}^2 R_{11} R_{22}R_{33} ,
\end{aligned}
\end{equation}

Substituting matirx \(\mathbf{P}\) into Eq. (\ref{UV}), we can obtain \( \mathbf{U} \) and \( \mathbf{V} \) satisfied by the new solution \( \mathbf{P} \). Then, we can verify that they satisfy Eq. (\ref{E1}). Furthermore, through Eqs. (\ref{E2}-\ref{E3}), we can obtain the coefficients of the common - part of the new solution, as follows: 
\begin{equation}\label{egamma}
    e^{2\gamma} = -C_0 \mu_3^2 ( \frac{(m_1^\dagger \mathbf{P_0} m_1)^2 \rho^2 \mu_{12}^2 + (m_1^\dagger \Lambda m_1)^2 R_{12}^2}{R_{33}R_{22}R_{11}}),
\end{equation}
where \( C_0 \) is an arbitrary constant.

there are two arbitrary complex numbers and their complex conjugates in the expression, which is still relatively complex for us. So we make the following transformation:
\begin{equation}
\begin{aligned}
    a_1 &= \frac{\overline{c_1}}{\overline{c_2}},\quad a_2 = \frac{c_2 }{c_1}, \quad a_3 = c_1\overline{c_2},\\
\end{aligned}     
\end{equation}

Then, substituting \( \psi \) from Eq. (\ref{f1}) into Eq. (\ref{cosettran}), we can obtain the expression for \( w \) and write it as the Weyl-Papapetrou metric: 
\begin{equation}\label{wp2}
\begin{aligned}
ds^2 &= -\mu \bar{\mu} \mu_3 \frac{\tilde{A}}{\tilde{B}}(dt - \frac{\mu_{12}}{\mu \bar{\mu}}  \frac{\tilde{C}}{\tilde{A}} d\phi)^2 \\
+ &C_0 \frac{\tilde{B}}{\tilde{H}} (dz^2 + d\rho^2) + \frac{\rho^2 \tilde{B}}{\mu \bar{\mu} \mu_3\tilde{A}}d\phi^2,
\end{aligned}
\end{equation}
where
\begin{equation}
\begin{aligned}
\tilde{A} &= -\mu_{12}^2 \left( \mu_{23}^2 \mu_{13}^2 a_1 + a_2 \mu_3^2 \right)^2 \rho^2\\ 
&+\mu_3^2 R_{12}^2 \left( -\mu_{23}^2 + \mu_{13}^2 a_1 a_2 \right)^2\\
\tilde{B} &=  \mu_{23}^4 \left( \rho^4 \mu_{12}^2 \mu_{13}^4 a_1^2 + \mu^2 \mu_3^2 R_{12}^2 \right)\\
        &- 2 a_1 a_2 \mu\overline{\mu} \mu_3^2  \mu_{13}^2 \mu_{23}^2 R_{11} R_{22} \\
        &+ \overline{\mu}^2 \mu_3^2 \left( \mu_{13}^4 R_{12}^2 a_1^2 a_2^2 + \mu^2 \mu_3^2 \mu_{12}^2 a_2^2 \right)\\
\tilde{C} &= R_{12}\left(R_{22}a_1^2\mu_{23}^2\mu_{13}^4a_2\rho^2 - R_{22}\mu_{23}^2a_2\mu^2\mu_3^2 \right.\\&\left. +\mu_{13}^2R_{11}a_2^2\mu_3^2\overline{\mu}^2a_1 - \mu_{13}^2R_{11}\rho^2\mu_{23}^4a_1\right)
\\
\tilde{H}&=\frac{\mu\overline{\mu}\mu_3\mu_{12}^2 \mu_{13}^2 \mu_{23}^2 R_{11} R_{22} R_{33} }{a_3^2},
\end{aligned}
\end{equation}
this is a three - soliton form of the accelerating Kerr metric, which is similar to the three - soliton form of the accelerating Kerr - NUT metric in Ref. \cite{ref14}. 

Next, we need to determine the coordinate transformation and the position of the soliton. For the line element (\ref{wp2}), we perform the following coordinate transformation: 
\begin{equation}\label{coortran}
\begin{aligned}
         &t\mapsto \sqrt{\alpha} \cdot \frac{t - 2 a \varphi}{\sqrt{1 + \alpha^2 a^2}},\quad \rho = \frac{\sqrt{QP}}{(1 + \alpha r x)^2}\\
        &\varphi \mapsto -\frac{1}{\sqrt{\alpha}} \cdot \frac{\alpha^2 a t + \left( 1 - \alpha^2 a^2 \right) \varphi}{\sqrt{1 + \alpha^2 a^2}}\\
        & z = \frac{(x + \alpha r)[(r - m)(1 + m \alpha x) + \sigma^2 \alpha x]}{(1 + \alpha r x)^2}\\
\end{aligned},
\end{equation}
the trajectory of each soliton \( k \) is
\begin{equation}\label{muk}
    \mu_k(\rho, z) = \omega_k - z \pm \left[ (\omega_k - z)^2 + \rho^2 \right]^{\frac{1}{2}},
\end{equation}
here, \( w_k \) is an arbitrary complex number.

Substitute (\ref{coortran}) into (\ref{muk}), and take the positions of the solitons as follows: 
\begin{equation}\label{mulocal}
    w = -\sigma, \quad \bar{w} = \sigma, \quad w_3 = \frac{1 - \alpha^2 (m^2 - \sigma^2)}{2\alpha},
\end{equation}
we can obtain the expression of the soliton as follows: 
\begin{equation}\label{muexpr}
    \begin{aligned}
        \mu &= \frac{(r-m +  \sigma )(1+ \alpha m x + x \alpha \sigma   )(1-\alpha r )(1 - x)}{(1 + \alpha r x)^2}\\
        \bar{\mu} &= \frac{(r  - m - \sigma)(1 + \alpha m x -  x \alpha \sigma)(1-\alpha r )(1 - x)}{(1 + \alpha r x)^2}\\
        \mu_3 &= \frac{(1 - \alpha^2 r^2)\left((1 + \alpha m x)^2 - \alpha^2 \sigma^2 x^2\right)}{\alpha (1 + \alpha r x)^2}
    \end{aligned},
\end{equation}
among them, \(\alpha\), \(a\), and \(m\) represent the acceleration parameter, the rotation parameter, and the mass parameter, respectively. And \( \sigma^2 = m^2 - a^2 \leq 0\).

We take the arbitrary parameters as:
\begin{equation}\label{a1a2c0}
    \begin{aligned}
    a_1 &= \frac{(\sigma-m ) \alpha}{a \left( 1 + \alpha m -\alpha \sigma \right)^2} ,\\
    a_2 &= \frac{(\sigma-m ) \left( 1 + \alpha m + \alpha \sigma \right)^2}{a \alpha},\\
    a_3 &=\frac{\alpha}{4\sigma ((1+m\alpha)^2-\alpha^2\sigma^2)}\\
    C_0 &= 4 \frac{(m + \sigma)^2 \left( 1 - (m - \sigma)^2 \alpha^2 \right)^2 }{\alpha^3 \left( 1 + \alpha^2 a^2 \right)},
    \end{aligned}
\end{equation}
substitute (\ref{coortran}), (\ref{muexpr}), and (\ref{a1a2c0}) into (\ref{wp2}), and perform analytic continuation on the real numbers \( a \) and \( m \) to make them lie on the imaginary axis. Here, \( x = \cos(\theta) \). Then we can obtain the common accelerating Kerr metric:
\begin{equation}
\begin{aligned}
    \mathrm{d}s^2& = \frac{1}{\Omega^2} \Bigg\{ -\frac{Q}{\varrho^2} \left( \mathrm{d}t - a \sin^2 \theta \mathrm{d}\varphi \right)^2 + \frac{\varrho^2}{Q} \mathrm{d}r^2  \\& + \frac{\varrho^2}{P} \mathrm{d}\theta^2 + \frac{P}{\varrho^2} \sin^2 \theta \left[ a \mathrm{d}t - \left( r^2 + a^2 \right) \mathrm{d}\varphi \right]^2 \Bigg\},
\end{aligned}
\end{equation}
where,
\begin{equation}
    \begin{aligned}
        \Omega &= 1 + \alpha r \cos\theta, \quad \varrho^2 = r^2 + a^2 \cos^2 \theta,\\
         P &= 1 + 2 \alpha m \cos\theta + \alpha^2  a^2  \cos^2 \theta,\\
          Q &= \left( a^2 - 2 m r + r^2 \right) \left( 1 - \alpha^2 r^2 \right).
    \end{aligned}
\end{equation}

Where \(m,a,and~ \alpha\) denote the mass parameter, rotation parameter, and acceleration parameter, respectively.

\section{Conclusion and Discussion}\label{sec4}

In this paper, by means of the reduction conditions of the nonlinear \(\sigma\)-model for the vacuum Einstein equation, we simplify the steps of the inverse scattering technique. Taking the pure Rindler metric as the background solution, we add a pair of solitons that are complex conjugates of each other and obtain a solution in the form of three solitons in the nonlinear \(\sigma\)-model. Through appropriate parameter selection and coordinate transformation, we prove that this solution is equivalent to the accelerating Kerr metric, which was not achieved in previous works. In previous works, Ref. \cite{ref18} constructed the possible modification conditions for black holes with acceleration parameters in four-dimensional supergravity. The stationary C-metric and its time-dependent solutions in the EMD theory have also been found \cite{ref19}. In Ref. \cite{ref20}, by adding two solitons to the flat spacetime as the background, the rotating soliton solution in EMDA was obtained, but the reduction problem in the nonlinear \(\sigma\)-model was not solved, instead, these conditions were merely treated as ansatzes.

In the future, we hope to use the inverse scattering technique to obtain accelerating solutions in the nonlinear \(\sigma\)-model of the EM theory and the EMDA theory. Ref. \cite{ref21} has already taken the accelerating Kerr metric as a seed at the low energy limit of the heterotic string theory and used the Hassan–Sen transformation to obtain a pair of rotating, charged, and accelerating black holes. Recently, taking the accelerating Kerr-Taub-NUT spacetime as the seed solution, the accelerating Kerr-Sen-Taub-NUT spacetime has been obtained through the Hassan-Sen transformation \cite{ref22}. These works provide a reference value for our future work. If we can solve the aforementioned reduction problem and select an appropriate seed solution, it will be possible to obtain the results we expect. 

\begin{acknowledgements}
 I thank my supervisor, Prof. S. Q. Wu, for entrusting me with this research topic, engaging in discussions with me on research details and offering valuable advice throughout the process. This work is supported by the National Natural Science Foundation of China (NSFC) under Grants No. 12375053.
 
\bigskip
\noindent\textbf{Data Availability Statement}~This manuscript has no associated data or the data will not be deposited. [Authors’ comment: There are no external data associated with the manuscript.]
\end{acknowledgements}

\appendix
\section{The reduction problem of symmetric spaces}\label{appendix}

In this appendix, we will derive the steps of the inverse scattering method in complex form for the SU(1,1)/U(1)-sigma model in detail and present a method to solve the reduction problem.

First, we need to introduce an exterior differential formulation that is independent of specific coordinate forms, which will greatly simplify our derivation steps. For matrices satisfying the SU(1,1)/U(1) group \cite{ref13}, we have:
\begin{equation}\label{Pcon}
    \mathbf{P} = \mathbf{P}^\dagger, \quad (\Lambda \mathbf{P})^2 = \mathbf{I}, \quad \det(\mathbf{P}) = 1, \quad \Lambda = \begin{pmatrix} 0 & i \\ -i & 0 \end{pmatrix},
\end{equation}
and we can define:
\begin{equation}
    \begin{aligned}
        W &= (-d \mathbf{P})\mathbf{P}^{-1}, \quad \Omega(\lambda, \rho, z) = \frac{\rho^2 W - \lambda \rho^* W}{\lambda^2 + \rho^2},\\
        D &= d - \left( \frac{\partial \Theta}{\partial \lambda} \right)^{-1} d\Theta \left( \frac{\partial}{\partial \lambda} \right).
    \end{aligned}
\end{equation}
Moreover, from the above definitions, we can readily obtain the following.
\begin{subequations}
    \begin{equation}\label{con1}
        \Omega^\dagger(\bar{\lambda}) = -\Lambda\Omega(\lambda)\Lambda,
    \end{equation}
    \begin{equation}\label{con2}
        \Omega(\lambda) =-\mathbf{P}\Lambda\Omega(\lambda)\mathbf{P}\Lambda,
    \end{equation}
    \begin{equation}\label{con3}
        \Omega(\tau) = W - \Omega(\lambda).
    \end{equation}
\end{subequations}
\( \bar{\lambda} \) denotes the complex conjugate of \( \lambda \),
where
\begin{equation}
\begin{aligned}
    \Theta(\lambda, \rho, z)& = \frac{\rho^2}{2\lambda} - \frac{\lambda}{2} - z, \quad \tau = -\frac{\rho^2}{\lambda}.
\end{aligned}
\end{equation}
the $*$ represents the Hodge dual, defined as: \( ^*d\rho = dz \), \( ^*dz = -d\rho \). This formulation has been rigorously verified in Ref. \cite{ref12}. Thus, we can simply express Eq. (\ref{LA}) as:
\begin{equation}\label{DPsi}
    D\Psi = -\Omega\Psi.
\end{equation}
Next, let us consider the following system of partial differential equations:
\begin{equation}
    D\left[ \Lambda \Psi(\lambda) \right] = -\Lambda \Omega(\lambda) \Psi(\lambda),
\end{equation}
taking the Hermitian conjugate of both sides of the system and substituting the complex conjugate of \( \lambda \), we can obtain:
\begin{equation}
    D\left[ \Psi^\dagger(\bar{\lambda}) \Lambda \right] = -\Psi^\dagger(\bar{\lambda}) \Omega^\dagger(\bar{\lambda}) \Lambda,
\end{equation}
substituting (\ref{con1}), we can get:
\begin{equation}
    D\left[ \Psi^\dagger(\bar{\lambda}) \Lambda \right] = \Psi^\dagger(\bar{\lambda}) \Lambda \Omega(\lambda),
\end{equation}
obviously, \( \Psi^\dagger(\bar{\lambda}) \Lambda \) and \( \Psi^{-1}(\lambda) \) satisfy the same system of partial differential equations, so we have:
\begin{equation}
    \Psi^\dagger(\bar{\lambda}) \Lambda = J_0 \Psi^{-1}(\lambda),
\end{equation}
let us consider the case when \( \lambda = 0 \). Given \( \Psi(0) = \mathbf{P} \), we can obtain \( J_0 = \Lambda \).
\begin{equation}
    \Psi^{-1}(\lambda)=\Lambda \Psi^\dagger(\bar{\lambda})\Lambda,
\end{equation}
next, consider \( \Psi = \chi(\lambda) \Psi_0 \), from which we can obtain a very important relation:
\begin{equation}\label{inchi=LchiL}
    \chi^{-1}(\lambda) = \Lambda\chi^\dagger(\bar{\lambda}) \Lambda.
\end{equation}

Substituting Eqs. (\ref{chi}), (\ref{inchi}), and (\ref{RS}) into the Eq. (\ref{inchi=LchiL}), and considering that the residue of the left-hand side of the equation at \( \lambda = \mu_k \) is not zero, to make this equation hold strictly, we must require that the right-hand side is also non-zero. Therefore, we have the following relations:
\begin{equation}\label{reduction1}
    p_k = \Lambda m_k, \quad \bar{\nu}_k = \mu_k
\end{equation}

It can be seen that this reduction condition reduces the number of arbitrary parameters we need by half, which greatly simplifies the difficulty of solving the new solution \( \mathbf{P} \). However, this condition only makes the matrix \( \mathbf{P} \) satisfy the Hermitian conjugate property. We still need additional constraint conditions to make the matrix \( \mathbf{P} \) fully satisfy the characteristic properties of the coset representation of \( \text{SU}(1,1)/\text{U}(1) \).

Next, let us consider a system of partial differential equations.
\begin{equation}
    \Lambda P^{-1} D\Psi(\lambda) = -\Lambda P^{-1} \Omega(\lambda) \Psi(\lambda),
\end{equation}
with the coordinate transformation \( \tau: \lambda \to -\frac{\rho^2}{\lambda} \), we can obtain:
\begin{equation}
\Lambda P^{-1} D\Psi(\tau) = -\Lambda  P^{-1} \Omega(\tau) \Psi(\tau),
\end{equation}
substituting in Eqs. (\ref{con2}) and (\ref{con3}), we can obtain:
\begin{equation}
    D\left[ \Lambda P^{-1} \Psi(\tau) \right] = -\Omega(\lambda) \left[ \Lambda P^{-1} \Psi(\tau) \right].
\end{equation}

It can be seen that this system of equations satisfies the same system of partial differential equations as \( \Psi(\lambda) \). The remaining steps are similar to the first reduction condition. To avoid verbosity, we directly present the final result.
\begin{equation}\label{P=chip0inchi}
    P = \chi(\tau) P_0 \Lambda \chi^{-1}(\lambda) \Lambda.
\end{equation}
It can be seen that the pole problem on the right-hand side of the equation in the complex plane is only related to \( \chi \) and \(\chi^{-1}\). So, consider that this expression should have exactly the same residues at \( \lambda = \mu_k \) and \( \lambda = \nu_k \) as the condition \( \chi(\lambda)\chi^{-1}(\lambda) = \mathbf{I} \). Therefore, we can obtain:
\begin{equation}\label{reduction2}
\begin{aligned}
    &\mu_i \mu_{i+1} = \nu_i \nu_{i+1} = -\rho^2, \quad (i = 1, 2, \dots, N-1),\\
   &m_{i+1} = \Lambda P_0 m_i, \quad p_{i+1} = P_0 \Lambda p_i
\end{aligned}
\end{equation}
It can be seen that this reduction condition reduces the number of arbitrary parameters by half again. For the \( \text{SU}(1,1)/\text{U}(1) \) group, we only need to consider two arbitrary complex constants. However, this condition only holds when the number of added solitons is even and cannot be satisfied for an odd number of solitons.

The reduction conditions for the SU(1,1)/U(1) sigma-model are (\ref{reduction1}) and (\ref{reduction2}). We believe that these two reduction conditions have minimized the parameters of this model. However, two arbitrary complex parameters (i.e., four arbitrary real parameters) are still required, and solving these parameters remains quite difficult. Especially when generalized to cases with an electromagnetic field, the parameters become more numerous and complex. We have also considered the reduction conditions in the EMDA coset representation \cite{ref20}, i.e., those satisfying the coset representation of \(\text{Sp}(4,\mathbb{R})/\text{U}(2)\) .

When considering the reduction conditions only in the real number field, (\ref{reduction1}) and (\ref{reduction2}) are contradictory, and only one of them can hold at a time. Of course, the actual equation will be modified to some extent.


\end{document}